\documentclass{optica-article}

\journal{opticajournal} 

\articletype{Research Article}

\usepackage{lineno}
\usepackage{amsmath}
\usepackage{graphicx}
\usepackage{romannum}
\usepackage{orcidlink}
\usepackage{comment}
\usepackage{gensymb}
\linenumbers 
\usepackage{fancyhdr,lipsum}

\begin{document}
\nolinenumbers

\title{Ultrafast Laser-Fabricated Fluoride Glass Waveguides with Exceptionally High Positive Refractive Index Change for Mid-Infrared Integrated Optics}

\author{T Toney Fernandez,\authormark{1,*} 
    	\orcidlink{0000-0002-6161-6496}
    Yongsop Hwang, \authormark{1}
    	\orcidlink{0000-0003-2283-9475}
    H Mahmodi \authormark{2}
    D E Otten,\authormark{1} 
    L Plenecassagne,\authormark{3} 
    S Cozic,\authormark{4} 
    S Gross, \authormark{5}
    \orcidlink{0000-0001-5130-183X} 
    I Kabakova,\authormark{2,7}
    M Withford, \authormark{6} 
    M Poulain,\authormark{4} 
    A Fuerbach, \authormark{6} 
    D G Lancaster\authormark{1,7}
    	\orcidlink{0000-0002-1299-8600}}

\address{\authormark{1}STEM, University of South Australia, Mawson Lakes Campus, South Australia - 5095\\
\authormark{2}School of Mathematical and Physical Sciences, University of Technology Sydney, Ultimo, NSW 2007, Australia\\
\authormark{3} Engineering School of 
University Paris-Saclay, Orsay, Essone, France\\
\authormark{4}Le Verre Fluoré, 1 Rue Gabriel Voisin - Campus KerLann, 35170, Bruz, Brittany, France\\
\authormark{5}School of Engineering, Macquarie University, NSW, 2109, Australia\\
\authormark{6}School of Mathematical and Physical Sciences, Macquarie University, NSW, 2109, Australia\\
\authormark{7}Australian Research Council COMBS Centre of Excellence}

\email{\authormark{*}toney.teddyfernandez@unisa.edu.au} 


\begin{abstract*} 
This study presents the successful fabrication of waveguides with a high positive refractive index change exceeding 0.02 in rare earth-doped fluoride glass, marking a major advancement in integrated optical components for visible to mid-infrared applications. By overcoming persistent challenges in mid-infrared direct-write photonics, this research enables the development of waveguides with high refractive index contrast and mode tailoring in optical substrates, supporting the miniaturization of optical devices. The investigation reveals that the exceptionally high index change results from material densification, driven primarily by the migration of barium within the glass composition. With low propagation losses (~0.21 dB/cm) and a highly customizable V-number over a broad wavelength range from visible to mid-infrared, these waveguides hold significant promise for chip laser technologies and the development of advanced optical devices for sensing and spectroscopy.

\end{abstract*}

\section{Introduction}


Fluoride glass is one of the few dielectric materials that exhibit broad transparency, spanning from the ultraviolet (UV) to the mid-infrared (MIR) range — an optical characteristic that sets it apart from traditional silica~\cite{Jackson2024, Maria2024}. \textcolor{black}{Unlike silica, which has limited transparency in the MIR region~\cite{Moore22}, fluoride glass enables the development of advanced photonic devices across a wider spectrum, making it an invaluable material for applications requiring extensive wavelength coverage~\cite{Hu2019}. Moreover, while rare-earth ions offer numerous visible transitions, these are typically suppressed in high-phonon energy hosts like silica due to non-radiative relaxation processes. Silica has a high phonon energy of approximately 1100 cm$^{-1}$, which significantly increases the likelihood of quenching lasing transitions. In contrast, fluoride glass, with much lower phonon energy (~500 cm$^{-1}$), offers reduced non-radiative losses, thereby facilitating efficient lasing at visible and infrared wavelengths.} However, despite the first report of fluoride glass waveguide fabrication by Ko et al.~\cite{Ko1990} in 1990, the development of high-performance waveguide platforms for visible to mid-infrared integrated optic circuits has been overdue for more than three decades. In their work, they reported an ion exchanged waveguide  with a large index change of $\approx 6 \times 10^{-2}$ through exchange of fluorine in glass with gaseous chlorine. \textcolor{black}{The large induced index was explained by the increase in density and change in polarizability($\alpha$, Polarizability refers to the ability of atoms or chemical groups to have their electron cloud distorted by an external electric field. This deformation influences how particles interact with electromagnetic waves, contributing to the refractive index across a range of frequencies) between $F^{-}$ ($m$=18.9984 g/mol, ionic radii $r$=1.36$\textup{~\AA}$, $\alpha$= 0.76-1.04~$\text{\AA}^3$) and $Cl^{-}$ ($m$=35.453 g/mol, $r$=1.81$\textup{~\AA}$, $\alpha$= 2.96-3.66~$\text{\AA}^3$) ions.} $OH^{-}$ and $OD^{-}$ were also explored for ion exchange, where D stands for Deuterium. The index changes obtained were exceptionally high in the order of 10$^{-2}$, but the production of channel waveguides was difficult due to a detrimental chemical reaction between fluoride glass and the deposited metal layer used as mask. Additionally, the process had a slow exchange rate requiring $\approx$8-10 hours to create a waveguide thickness of  $\approx$6-8 \textmu  m~\cite{LUCAS20013205, JOSSE1997152, JOSSE19971139, HAQUIN2003460}. Cationic exchanges with $Li^{+}$, $Na^{+}$ and $K^{+}$ were also explored in fluoride glasses with limited success due to severe devitrification and fast corrosion of the substrates ~\cite{FOGRET199679}. Physical vapor deposition of fluoride glass on fluoride crystal substrates were also proposed but the procedure was challenging due to moisture sensitivity, stress buildup and losses~\cite{BOULARD200072}. \textcolor{black}{In another unique study, Cho et al. \cite{CHO} demonstrated permanent structural transformations in a multicomponent glass composed of $ZrF_4, BaF_2, LaF_3, AlF_3,$ and $NaF$ (ZBLAN), induced by a self-channeled plasma filament.} When an ultrashort laser pulse of 800 nm wavelength and 110 fs duration was focused inside ZBLAN glass, filamentation occurred due to the dynamic balance between Kerr self-focusing and defocusing effects caused by the electron plasma generated during the ionization process. The photoinduced refractive index modification extended over a length of 10-15 mm, with diameters ranging from 5 to 8 \textmu m. The refractive index change was measured to be as high as 1.3 $\times~10^{-2}$. There have been no further reports since then, possibly due to the limitation that only straight-line geometries could be induced, making precise control over the V-number and waveguide length challenging. \par
With the demonstration of femtosecond laser micromachining in silicate glasses in the mid 90's, fluoride glasses were also trialled, but efforts to produce high positive index change ($>10^{-2}$) fluoride glass waveguides were not achieved. The initial report by Miura $\textit{et.al}$ demonstrated tunability of waveguide width and index change with different pulse widths and multiple scan passes. A positive index change of $3 \times 10^{-3}$ was reported~\cite{MIURA1999212}. Depressed cladding structures, where negative index changes are stacked concentrically to create a cladding of lower refractive index than the bulk were also successful, but limited to large mode area waveguides~\cite{Lancaster:11}. Typical index contrasts obtained for these structures ranged between $- 1 \times 10^{-3}$ to $- 3 \times 10^{-3}$~\cite{Gross:13,Bernier:07}.\par

In 2013 continued efforts to produce smooth high positive index change were resumed in an exhaustive search across a broad parameter space of repetition rates (1--250 kHz), fluences (5 kJ/cm$^2$ -- 100 MJ/cm$^2$) and feedrates between 50--5000 \textmu m/s, reporting a maximum of $1.25 \times 10^{-3}$ by Berube $\textit{et.al}$~\cite{Berube13}. In 2014, a modified fluoroborate glass containing $WO_3$ was used to report a positive index change value of $\approx 5 \times 10^{-3}$~\cite{Ledemi14}. Later, up to $6.4\times 10^{-3}$ positive index change was reported in a modified fluoride glass that contained a second glass former (HfF$_4$). These waveguides were used to propagate a single mode at 3.1 $\mu m$ wavelength~\cite{Fernandez2022}.\par
\begin{figure}[t]
    \centering
    \includegraphics[trim={0 0 0 0},width=10 cm]{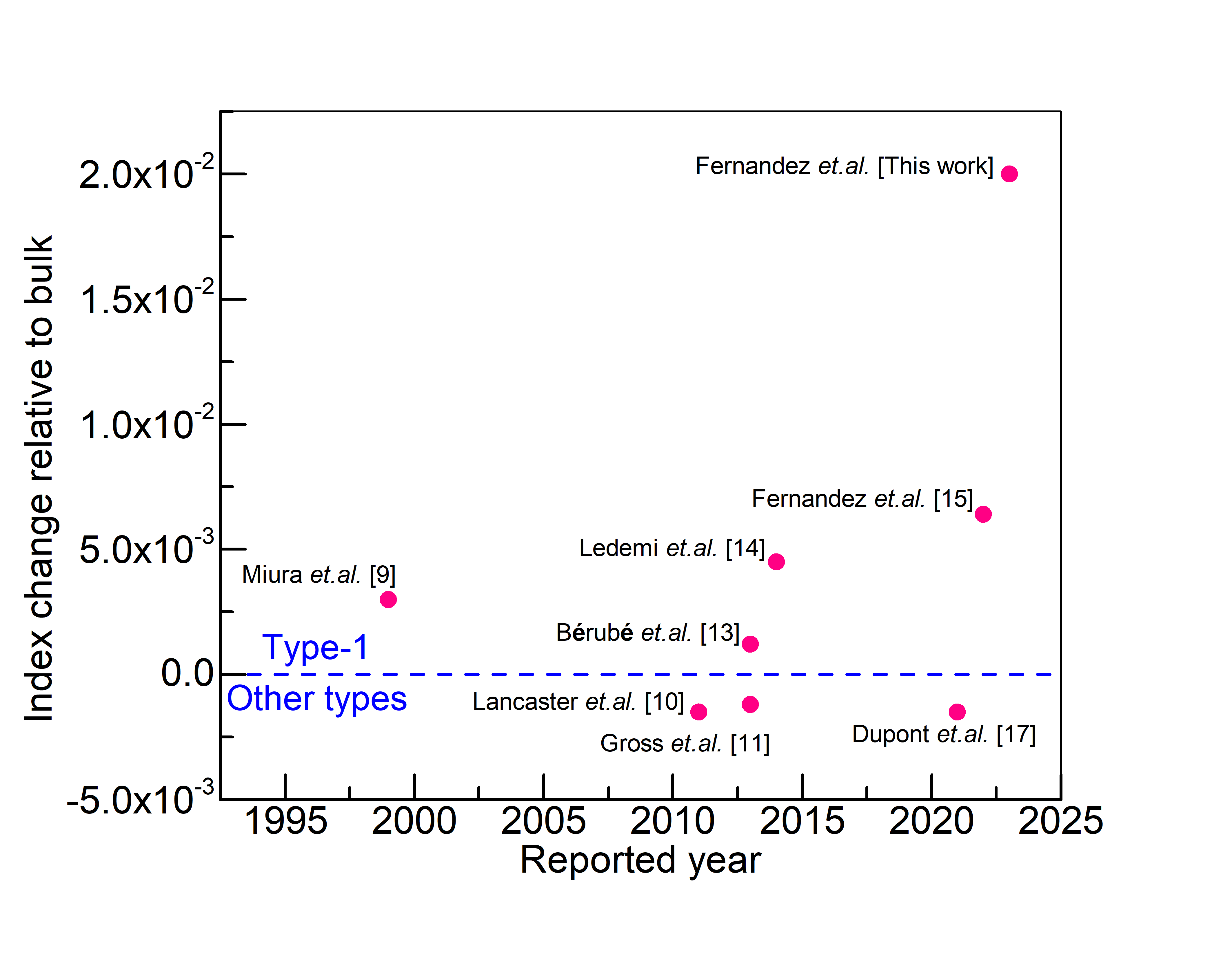}  
        \caption{Magnitude of index change reported in fluoride dielectrics by the fs-laser 3D direct-write technique.}
    \label{ZBLAN_Timeline}
    \end{figure}

\textcolor{black}{A comprehensive discussion on the importance of achieving a high positive index change in fluoride glasses, including a detailed analysis of the dependence of mid-infrared wavelengths, refractive index change, and waveguide dimensions required for single-mode operation, can be found in ~\cite{fernandez2024}.} In the work presented here, a record-high positive index change waveguide with low loss in a modified fluoride glass is reported for the first time. A consolidated time line of refractive index change obtained using the femtosecond laser direct-write technique are depicted in Fig.~\ref{ZBLAN_Timeline}. The reported propagation loss for those waveguides along with their inscription parameters are tabulated in Table~\ref{table:1} 

\begin{table}[h!]
\small
\centering
\begin{tabular}{c c c } 
 \hline
 Ref. & $\lambda$, Pulse width, Rep.rate, Obj.NA, feedrate, WG Type & Prop.loss @ $\lambda$\\ 
 \hline
Lancaster et.al~\cite{Lancaster:11} & 800 nm, 50 fs, 5.1 MHz, 1.25 NA, 1000mm/min,, T-\Romannum{2} & \textbf{0.22~dB/cm @ 1.9~\textmu m}\\
 Gross et.al~\cite{Gross:13} & 800 nm, 50 fs, 5.1 MHz, 1.25 NA, 1000mm/min,, T-\Romannum{2} & \textbf{0.29~dB/cm @ 4~\textmu m}\\ 
 Dupont et.al~\cite{Dupont21} & 515 nm, 270 fs, 3.15 MHz, 0.8 NA, 5mm/s, T-\Romannum{2} & \textbf{0.97~dB/cm @ 2.85~\textmu m}\\
 Berube et.al~\cite{Berube13} & 792 nm, 70 fs, 100-250 kHz, 0.1 NA, 20 \textmu m/s, T-\Romannum{1} & \textbf{0.4 to 2~dB/cm @ 0.633~\textmu m}\\
 Fernandez et.al~\cite{ Fernandez2022} & 1030 nm, 240 fs, 5 kHz, 0.6 NA, 0.04 mm/s,, T-\Romannum{1} & \textbf{1.37~dB/cm @ 3.1~\textmu m}\\
 \textcolor{red}{This work} & 1030 nm, 240 fs, 50 kHz, 0.6 NA, 0.04 mm/s,T-\Romannum{1} & \textbf{0.21~dB/cm @ 1.625~\textmu m}\\
 \hline
\end{tabular}
\caption{Reported propagation loss values of positive index and depressed index cladding fluoride glass waveguides. }
\label{table:1}
\end{table}

\section{Materials and Methods} \label{section2}
The fluoride glass composition used in the past for fs-laser direct-write waveguide inscription, is the most common composition  found in the literature: $53 ZrF_4  ~20 BaF_2  ~3 LaF_3  ~4 AlF_3  ~20 NaF$ (mol\%)~\cite{Gross:13}. \textcolor{black}{Within this composition $ZrF_4$ is a glass former (compounds that can form a glassy structure by creating a continuous, disordered network of bonds, resulting in an amorphous, non-crystalline solid.) whereas $BaF_2, LaF_3 and NaF$ are network modifiers (compounds that, when added to glass formers, disrupt the continuous network of bonds, altering the optical, structural and mechanical properties of the glass without forming a network on their own.) and finally $AlF_3$ is the conditional glass former or otherwise known as intermediate (compounds that can either behave as glass formers or modifiers, depending on the composition and conditions of the glass system).}  To date, the formation of optical waveguides in fluoride glasses by ultrafast laser inscription has not been associated with the migration of ions/elements leading to material densification. In the past, we hybridized the zirconium glass former with hafnium (Hf), successfully increasing the positive index contrast to up to $6.4 \times 10^{-3}$~\cite{Fernandez2022}. This attempt was based on the previous approach of producing positive index changes in amorphous dielectrics which otherwise would produce only negative index change or weak positive index changes,  such as phosphates (matrix hybridized with lanthanum to produce waveguides with $\Delta n_{positive} = 1.5 \times 10^{-2}$)~\cite{Fernandez13}, tellurite (with phosphate, $\Delta n_{positive} = 3.5 \times 10^{-3}$)~\cite{Fernandez14} and gallium lanthanum sulphide (with oxides, $\Delta n_{positive} = 4.9 \times 10^{-3}$)~\cite{Gretzinger:20}. Though, the hybridization attempt of ZBLAN with Hf did not show any elemental migration and the increase in index was purely based on the change in polarizability of constituent molecules due to laser pulse induced bond breaking. It is suspected that the absence of elemental migration in the fluoride glass composition may be attributed to the mixed modifier effect, as these compositions typically include at least three different modifiers: barium, lanthanum, and sodium (with $AlF_3$ acting primarily as a conditional glass former). Mixed modifier typically means that different types of modifier ions (or elements) are competing/interacting in a way that affects their ability to occupy certain sites within the glass network. Hence, one modifier might have trouble occupying the same site or interacting in the same way as another modifier that was previously occupying that site~\cite{Dyre_2009}. This competition can lead to the lack of elemental migration in fluoride glass composition upon fs-laser irradiation. For the first time, this is addressed by revising the amount of network modifiers rather than network formers, specifically by removing $LaF_3$ and leaving the glass with a single multivalent glass modifier. Since the stability of zirconium-based fluoride glass primarily depends on the combined presence of $AlF_3$, NaF and $BaF_2$ ~\cite{Ohsawa}, the removal of $LaF_3$ should not impact its stability ~\cite{POULAIN2022}. A study involving the gradual modification of a ZBAN composition, $52 ZrF_4\-~24 BaF_2\-~4AlF_3\-~205\-~NaF (mol\%)$ by increasing the $LaF_3$ content at the expense of NaF, reported an increase in both the glass transition and crystallization temperatures, along with a decrease in the thermal expansion coefficient.  Thermal and optical comparison of a ZBLAN and ZBAN glass composition is provided in Table~\ref{table:2} taken from ~\cite{POULAIN2022,SAAD}.

\begin{table}[h!]
\small
\centering
\begin{tabular}{c c c c c c c} 
 \hline
Glass & Tg ($^{\circ}$C) & Tx ($^{\circ}$C) & d (gcm$^{-3}$) & $n_D$ &  $\alpha$ (10$^{-7}$K$^{-1})$ & Transparency (\textmu m)\\ 
 \hline
 ZBLAN & 311--263 & 390--370 & 4.52 & 1.498 & 162-200 & 0.4 - 4.5 \\
 ZBAN  & 262--255 & 351--340 & 4.28 & 1.497 & 200 & 0.4 - 4.5 \\ 
 
 \hline
\end{tabular}
\caption{Optical and thermal properties of fluoride glass with and without lanthanum ~\cite{POULAIN2022, SAAD}. Tg and Tx are the glass transition and crystallization temperatures respectively. d is density, $n_D$ is the refractive index measured at 589 nm and $\alpha$ is the coefficient of thermal expansion. }
\label{table:2}
\end{table}

The final composition of the glass that was chosen to inscribe waveguides was: $51.3 ZrF_4\-~19 BaF_2\-~3AlF_3\-~19.45\-~NaF\-~0.5ErF_3\-~1.75YbF_3\-~5CeF_3$ (mol\%), where the rare-earth elements were doped to use the generated waveguides as gain-medium for an erbium chip laser. Such a composition will not produce an IR edge absorption shift affecting the transparency compared to a ZBLAN glass~\cite{MOYNIHAN198125}. The glasses were prepared using the conventional melt-quenching technique at Le Verre Fluore's industrial facility in Brittany, France.

\begin{figure}[t]
    \centering
    \includegraphics[trim={0 0 0 0},width=6 cm]{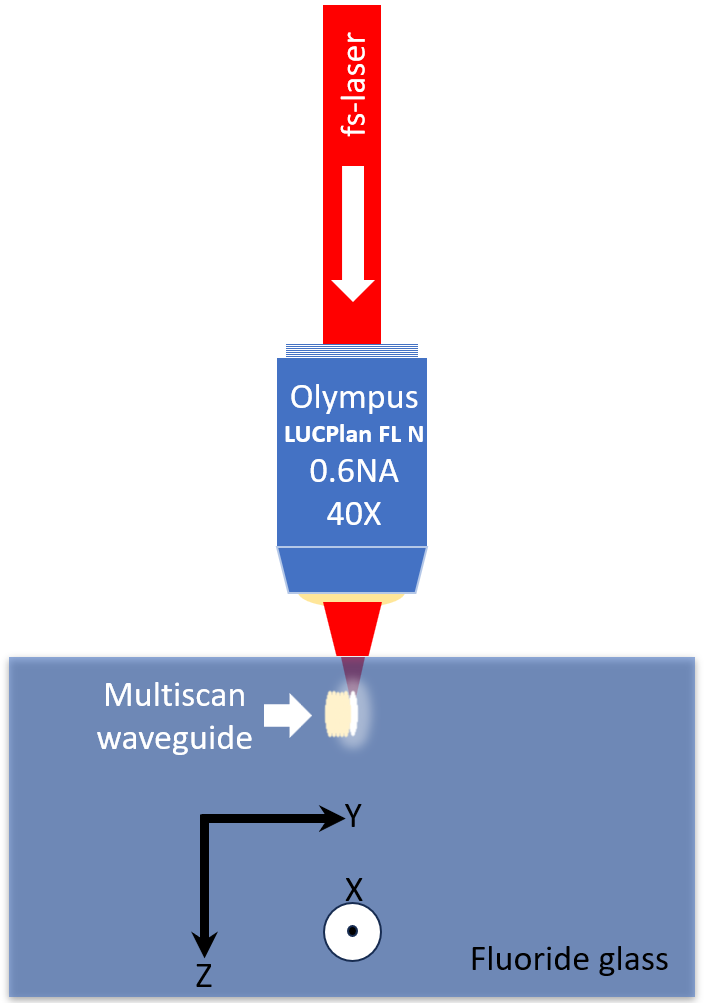}  
        \caption{\textcolor{black}{A simplified schematic of the fs-laser direct-write setup illustrates the X, Y, and Z axes for orientation. A single strand of the multiscan waveguide is created by translating the glass along the X-direction, while the complete waveguide is constructed by shifting (waveguide pitch) the sample along the Y-direction for each successive scan.}}
    \label{Setup}
    \end{figure}

\textcolor{black}{A Pharos femtosecond laser system (Light Conversion), operating at a central wavelength of 1030~nm, circularly polarized, with a pulse duration of 240~fs and a repetition rate of 50~kHz, was used to inscribe the waveguides.} As reported previously, lower repetition rates provide an optimised inscription window for multiscan waveguides in fluoride glasses~\cite{Bellouard04,Fernandez2022}. To account for the different glass composition, a parameter scan was conducted with repetition rates ranging from 1 to 250 kHz. It was found that also 50 kHz to be ideal for producing high positive index change waveguides in the amended composition. Introducing a precise amount of spherical aberration by detuning the coverslip compensation collar position of the focusing objective (Olympus, LUCPlan FL N, 0.6 NA, 40X) helped in maximising the achievable index change of the waveguides. \textcolor{black}{Spherical aberration assisted waveguide inscription is a successful technique~\cite{Fernandez2022, Fernandez2022_APLPh} to induce elemental migration due to thermal gradient of the focal spot within a dielectric medium. The underlying mechanism of this technique is explained both mathematically and geometrically in ~\cite{Song2011}, with further insights gained through laser-induced plasma imaging of the resulting thermal gradient, as detailed in ~\cite{Fernandez2015}.} For the current study, we found that a 500~\textmu m collar position was optimal for an inscription depth of 170~\textmu m. \textcolor{black}{A schematic of the fs-laser writing setup with the orientation of the X, Y, and Z axes is provided in Fig.~\ref{Setup}.}

\section{Results and discussion}
\begin{figure}[h]
    \centering
    \includegraphics[trim={0 0 0 0},width=9 cm]{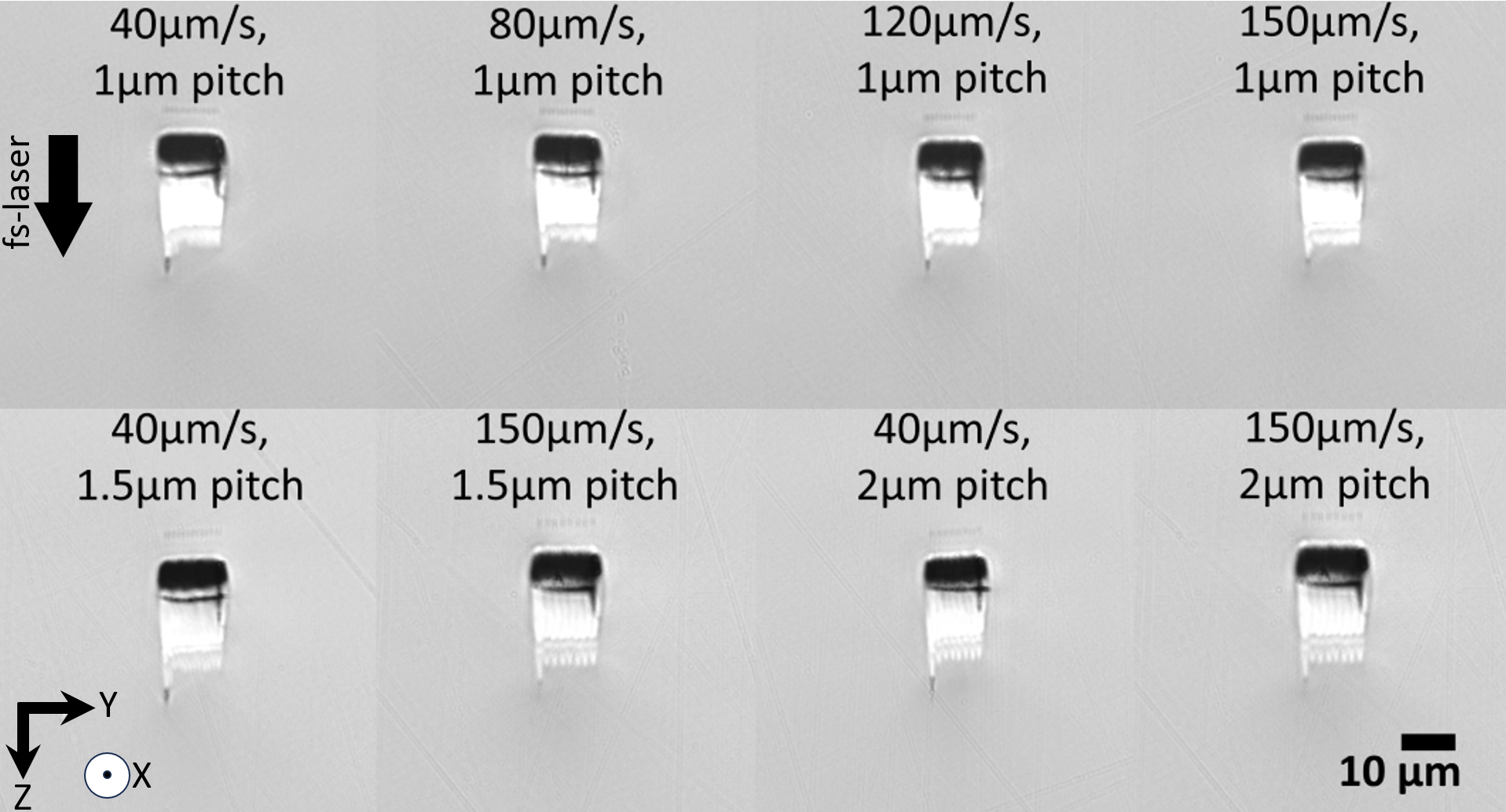}  
        \caption{(a) End-on differential interference contrast microscope images of waveguides which are nominally 12~\textmu m wide. The waveguides were inscribed using 50~kHz laser repetition rate and 4~\textmu J pulse energy measured before the objective. The feedrates and the lateral pitch between each track are provided on respective images. The laser was incident from the top and individual scans of each  multiscan waveguide were inscribed from left-to-right sequentially by feeding the sample in the same direction.}
    \label{DIC}
    \end{figure}

    \begin{figure}[h]
        \centering
    \includegraphics[trim={0 0 0 0},width=6.5 cm]{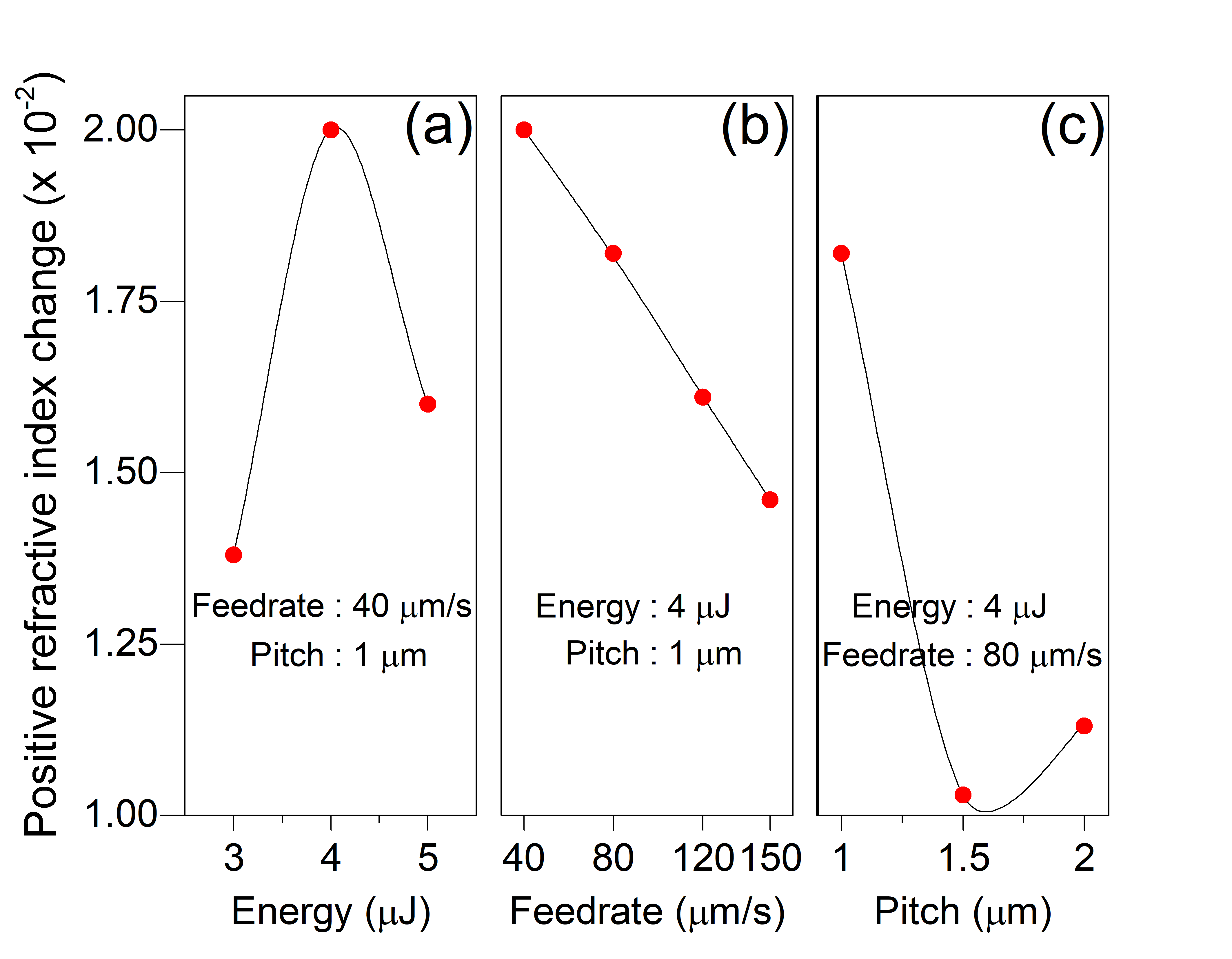} 
    \centering
    \includegraphics[trim={0 0 0 0},width=5 cm]{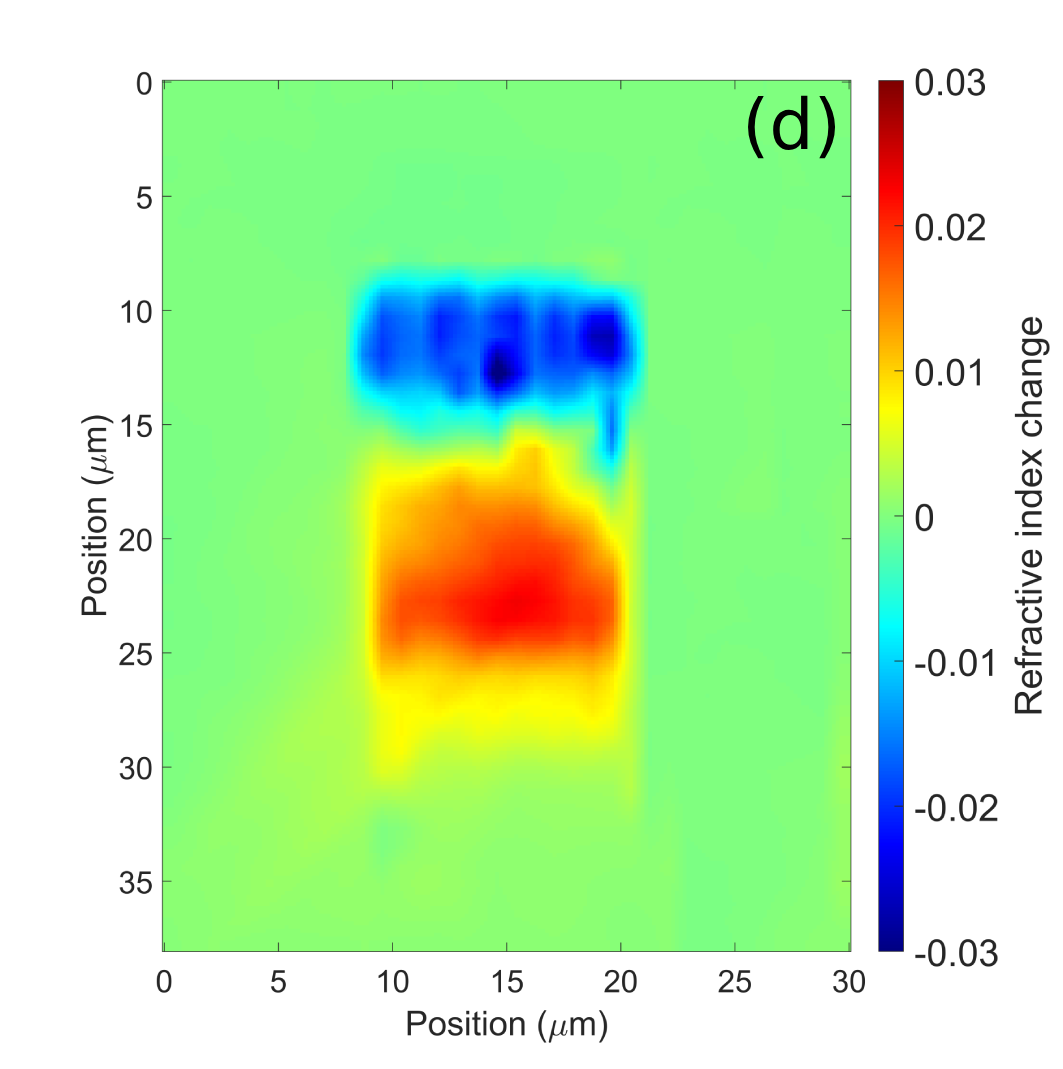}
            \caption{Measured peak positive index change as a function of (a)  pulse energy (fixed 40~\textmu m/s feedrate, 1~\textmu m pitch) , and (b) feedrate (fixed 4~\textmu J pulse energy, 1~\textmu m pitch),  (c) multiscan pitch (fixed 4~\textmu J pulse energy, 80~\textmu m/s feedrate). Solid lines are just guides for the eye. (d) Refractive index profile of a waveguide inscribed with an energy of 4~\textmu J, feedrate of 40~\textmu m/s and a pitch of 1~\textmu m. \textcolor{black}{Maximum positive index change = 2 $\times$ 10$^{-2}$ and maximum negative index change = $-$2.92 $\times$ 10$^{-2}$}}
    \label{Rinck}
    \end{figure}
Differential interference contrast (DIC) microscope images of the multiscan waveguides' cross-sections are shown in Fig.~\ref{DIC}. \textcolor{black}{The direction of fs-laser is shown in the figure and the axes are same as the schematic shown in Fig.~\ref{Setup}. A single laser scan across the sample produced a modification of positive index contrast that was approximately 1~\textmu m wide in the horizontal direction and 12-–14~\textmu m tall in the vertical dimension along the Z-axis.} 12~\textmu m wide waveguides were fabricated by laterally shifting each scan by preset pitches of 1, 1.5, 2 and 2.5~\textmu m in the horizontal direction. \textcolor{black}{The total number of scans was adjusted to maintain a nominal waveguide width of 12~\textmu m of which scans were carried out along the Y-axis sequentially and by feeding the sample in the same direction.} The best fabrication window with respect to morphology was provided using pulse energies ranging between 2--6~\textmu J, feedrates of 40--150~\textmu m/s and a multiscan pitch of 1~\textmu m which was the lowest pitch tested. Outside this window, the index contrast was either weak due to insufficient pulse energy or spatial overlap between modifications, or the structures exhibited mechanical fractures due to high pulse energies. Figure~\ref{DIC} shows that the waveguides have a negative index change (black regions) at the top and a positive index change on the bottom (white regions). This morphology is characteristic for multiscan waveguides~\cite{Fernandez2022, Fernandez2022_APLPh}.\\
The refractive index changes of the waveguides were measured using a Rinck refracted near-field profilometer at a wavelengths of 635 nm,  as shown in Fig~\ref{Rinck}. The peak positive refractive index changes are plotted in Figures~\ref{Rinck} (a-c) in dependence of the pulse energy, feedrate and multiscan pitch while keeping two out of the three parameters constant. \textcolor{black}{The index change shows a linear relationship when the feed rate is varied, compared to a quadratic relationship when energy and pitch are adjusted. This linearity can be attributed to the feed rate being the primary factor controlling the quench rate during waveguide formation (although the repetition rate can influence the quench rate, it remains constant in this case). Changing the laser energy only affects the quench temperature, and in this scenario, the impact on the quench rate is much smaller compared to doubling or tripling the feed rate. Adjusting the multiscan pitch influences the waveguide’s packing density, a parameter dependent on the material. Currently, a detailed discussion on the inflection points observed in Fig~\ref{Rinck}(a) and (c) is difficult due to the limited number of data points available.} Figure~\ref{Rinck} (d) presents a 2D refractive index profile of the waveguide fabricated with an energy of 4~\textmu J and a feedrate of 40~\textmu m/s. The positive refractive index change for this waveguide reached as high as 2 $\times$ 10$^{-2}$. The significant decrease in the refractive index at the top of the structure ($-$2.92 $\times$ 10$^{-2}$) enhances the effective index contrast experienced by a propagating mode.

\begin{figure}[t]
        \centering
        \includegraphics[trim={0 0 0 0},width=13 cm]{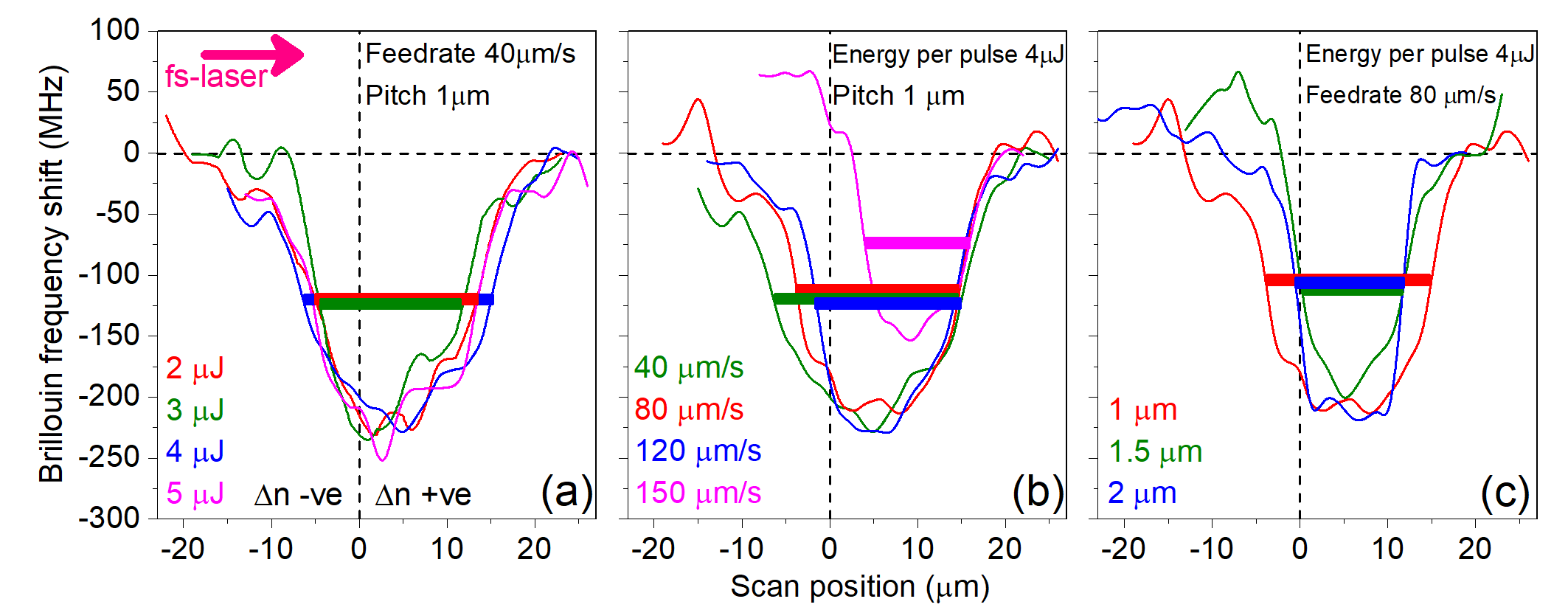}\\
        \includegraphics[trim={0 0 0 0},width=6 cm]{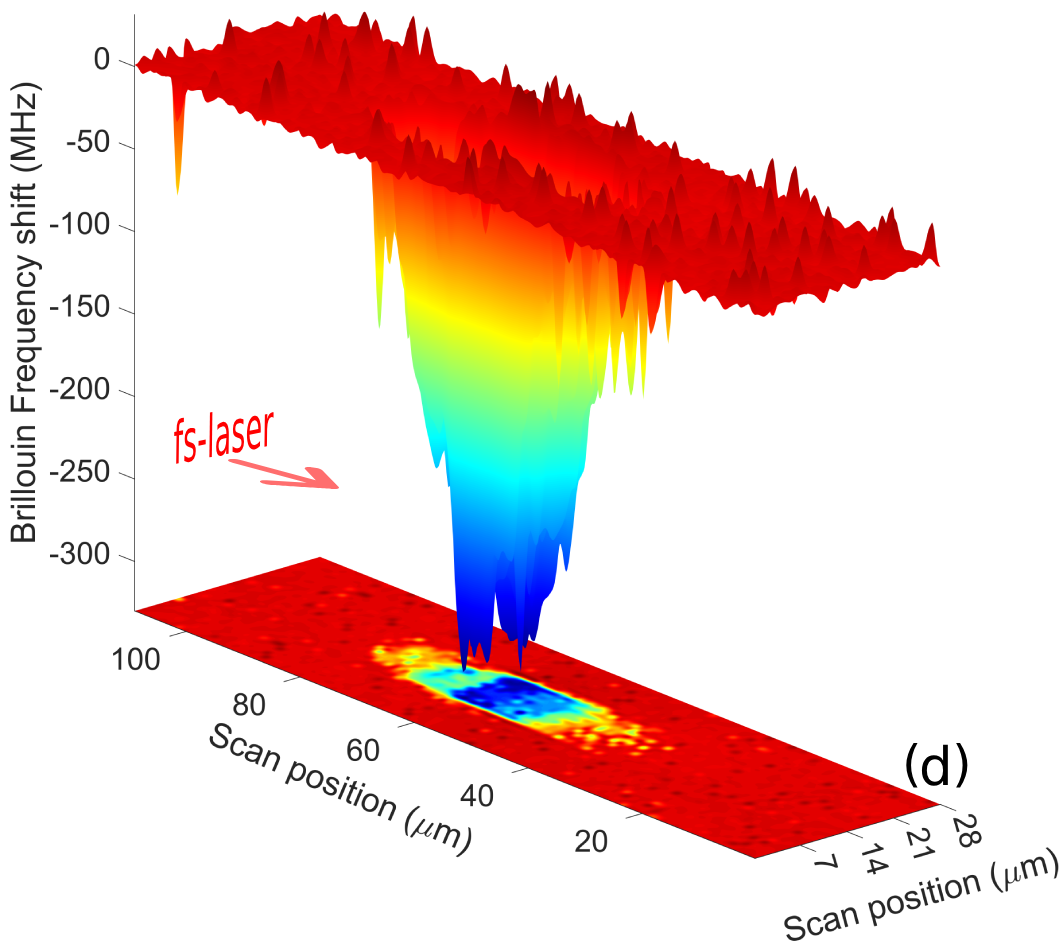} \includegraphics[trim={0 0 0 0},width=7 cm]{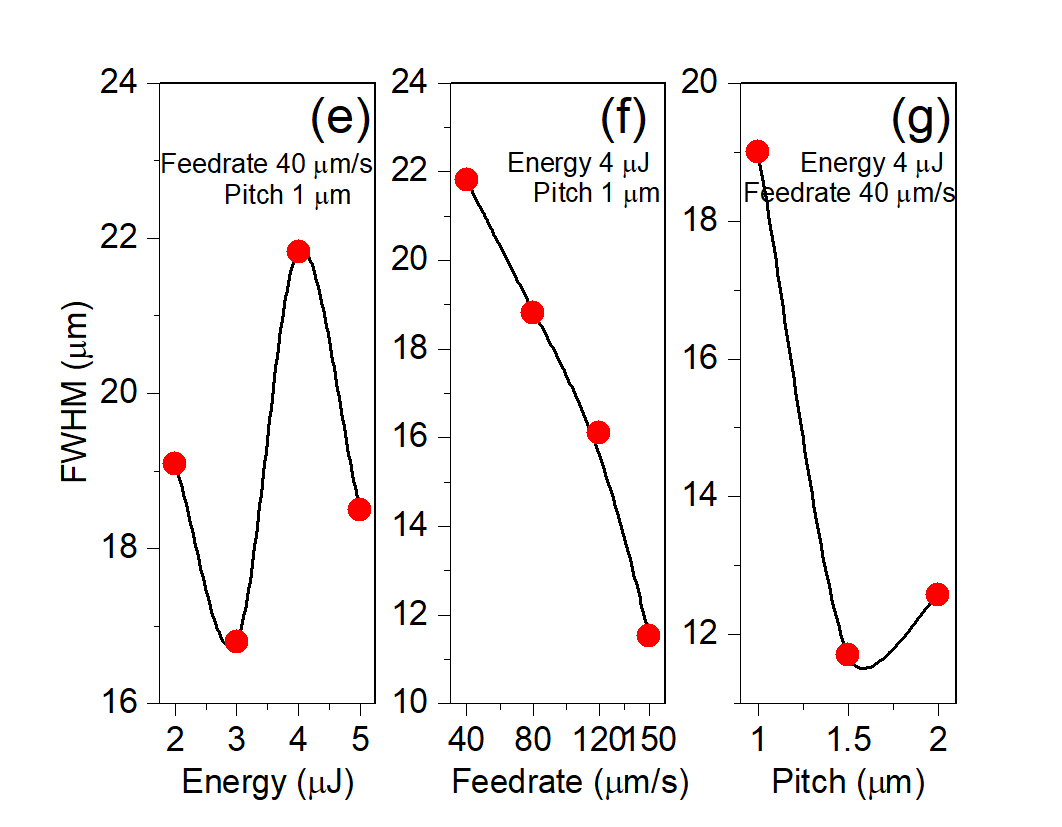}
    
            \caption{\textcolor{black}{(a) Spatial BFS map for a waveguide written with a pulse energy of 4~\textmu J, feedrate of 40~\textmu m/s and a pitch of 1~\textmu m. Measured FWHM of BFS along the whole laser induced structure as a function of (a)  pulse energy (fixed 40~\textmu m/s feedrate, 1~\textmu m pitch) , and (b) feedrate (fixed 4~\textmu J pulse energy, 1~\textmu m pitch),  (c) multiscan pitch (fixed 4~\textmu J pulse energy, 80~\textmu m/s feedrate). Solid lines are just guides for the eye.  (e-g) Line scans along the vertical direction of the waveguides, parallel to the inscription laser propagation direction as a function of (e) energy, (f) feedrate, and (g) pitch while keeping two out of the three parameters fixed. The fs-laser inscription direction is marked in (e) and the line scan is done along the z-axis shown in Fig.~\ref{Setup}}}
     \label{BFS}
    \end{figure}

This is the first time an index contrast of this magnitude is achieved in fluoride glass using ultrafast laser inscription. We have investigated its origin and traced it back to material densification which is clearly distinguishable in the Brillouin frequency shift (BFS) spectroscopy and back-scattered electron (BSE) microscopy. BFS were measured using a 660 nm single-frequency Cobolt Flamenco laser (H\"UBNER Photonics) through a confocal microscope (CM1, TableStable Ltd.) \cite{Fernandez2022}. \textcolor{black}{ Figure~\ref{BFS}(a-c) shows individual line scans along the direction of the inscription laser, obtained by varying the parameter space. The quadratic like behaviour and asymptotic tails is due to the induced strong thermal gradient from the aberration assisted laser inscription as detailed in section 2 and also in~\cite{Fernandez2015}. The x-axis, with 0~\textmu m, is aligned to the transition zone between the negative and positive refractive index changes. The total modified volume and BFS magnitudes suggest that altering the energy per pulse (Fig.~\ref{BFS}(a)), while holding other parameters constant, results in minimal changes. However, when the energy remains constant and the feedrate is reduced from 150 to 40~\textmu m/s (Fig.~\ref{BFS}(b)), the densified zone begins to expand. A shift from red to blue in the negative index change region is observed at the fastest tested feedrate of 150~\textmu m/s (represented by the magenta curve in Fig.~\ref{BFS}(b)). This could be due to the lower fluence, which reduces the formation of nano/micro voids, as previously reported~\cite{Fernandez2022}. Finally, varying the pitch, as illustrated in Fig.~\ref{BFS}(c), while keeping the other parameters constant, demonstrates that smaller pitches produce larger modification areas (broader bandwidth versus larger BFS), indicating that further pitch reduction could enhance the refractive index change. Figure~\ref{BFS}(d) shows the mapping of the Brillouin frequency shift across the laser-modified region for a waveguide inscribed using a pulse energy of 4~\textmu J, feedrate of 40~\textmu m/s and a pitch of 1~\textmu m, i.e. the waveguide shown in Fig~\ref{Rinck} (d). To clarify the different shifts discussed: the BFS refers to the conventional frequency shift (the anti-Stokes component) observed in bulk glass relative to the stimulating laser. However, due to laser irradiation, the modified zones exhibit small variations in BFS, referred to here as red or blue shifts relative to the unmodified bulk glass, i.e. a negative value of BFS in Figure~\ref{BFS}(d) refers to a smaller absolute BFS as compared to the bulk glass (hence referred to as red-shift). The data clearly indicates that the laser-induced structural changes have caused a red-shift in the Brillouin frequency compared to the unmodified bulk material, with the largest observed shift being approximately $-$260 MHz. This red-shift is novel, as previous reports have only documented blue-shifts in femtosecond laser-modified volumes, such as in hafnium modified fluoride glasses~\cite{Fernandez2022} and pure fused silica~\cite{Fernandez2022_APLPh}. According to the Brillouin frequency shift relation of $BFS=\frac{2n}{\lambda} \sqrt \frac{M}{\rho}$, where n is the refractive index, M is the longitudinal modulus, and $\rho$ is the density, a positive change in the refractive index accompanied by an increase in density can lead to either a blue or red shift in the Brillouin frequency. Atomic force microscope (AFM) measurements within the positive index change zones revealed no significant variation in M, suggesting that the observed red-shift in Brillouin frequency is primarily due to strong densification, i.e an increase in $\rho$. On the other hand, the red-shift observed within the negative index change zones, consistently reported in various other glasses like fused silica, borosilicates, and Z/HBLAN, is largely caused by the formation of a highly compressible open glass network including the presence of nano/micro voids. This results in a smaller longitudinal modulus and decreased refractive index which dominate any density changes, ultimately causing a negative BFS. Figure~\ref{BFS}(e-g) presents the spatial distance, within a waveguide, between two data points whose value is half of the BFS peak value. This is also represented as colored bands in Fig~\ref{BFS}(a-b) and hence the y-axis is measured in~\textmu m. This half-of-peak-value spacing (this is a spatial separation between those points where the BFS drops to half the peak, rather than a true FWHM) is relevant in waveguides as it corresponds to its normalized frequency parameter or otherwise known as V-number that characterizes the number of modes an optical waveguide can support. It is given by the formula: $V=\frac{2\pi a}{\lambda} NA$, where $a$ is the radius of the waveguide, $\lambda$ is the operating wavelength and NA is numerical aperture of the waveguide. Since NA reflects the refractive index change (NA$\approx$ $\sqrt{n_{waveguide}^2-n_{bulk}^2}$) and $a$ represent the spatial extent of the positive index change, we believe that plotting the half peak spacing provides a direct insight to the expected V-number of the waveguide from the laser parameters. It is interesting to note that the response mirrors the refractive index data shown in Fig.~\ref{Rinck}(a-c), as the index change also displays an inflection point at 4~\textmu J when feedrate and pitch are held constant, and at 1.5~\textmu m pitch when feedrate and energy are kept constant. It also indicates that the half peak spacing varies linearly when energy and pitch remain constant, closely resembling the trend in Fig.~\ref{Rinck}(b).}

\begin{figure}[t]
    \centering
    \includegraphics[trim={0 0 0 0},width=10 cm]{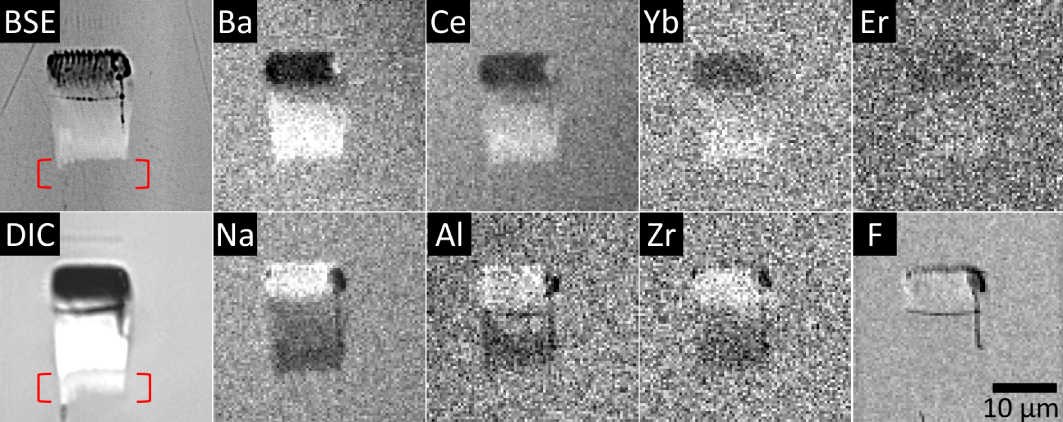}  
\caption{BSE, DIC and elemental maps of a waveguide inscribed with an energy of 4~\textmu J, feedrate of 40~\textmu m/s and a pitch of 1~\textmu m. Ba, Ce, Yb and Er is observed to migrate to the region of positive index change, while Na, Al and Zr migrate  to the negative index change region.  The red brackets in the BSE and DIC image indicate a zone of increased refractive index that is not associated with elemental migration, densification or rarefaction.}
    \label{EPMA}
    \end{figure}

BSE microscopy and electron probe microanalysis (EPMA) were carried out on a Cameca SXFive Electron Microprobe equipped with wavelength-dispersive spectrometers (WDS). BSE image (Fig.~\ref{EPMA}) obtained for waveguide inscribed with an energy of 4~\textmu J, feedrate of 40~\textmu m/s and a pitch of 1~\textmu m shows strong Z-contrast, where Z corresponds to atomic number. The negative index region reveals the presence of nano voids, a phenomenon previously documented in our studies on multiscan waveguides in both ZBLAN and fused silica~\cite{Fernandez2022, Fernandez2022_APLPh,Bhardwaj22}. Notably, a distinct horizontal string of nano voids was observed to be disconnected from the primary negative index zone. Furthermore, a strip of material located at the base of the waveguide, as indicated by red brackets in the differential interference contrast (DIC) image was conspicuously absent in both the back-scattered electron (BSE) image and the elemental mapping. We attribute this absence to a heat-induced diffusion or a change in polarizability, which manifests itself as an exclusive change in refractive index without any associated changes in density or composition. This behavior has been extensively elucidated in the context of fused silica~\cite{ Fernandez2022_APLPh}.
A comprehensive elemental mapping of the waveguides indicates that the positive index change is primarily driven by densification caused by the migration of Ba, Ce, Yb, and Er. Notably, Ba is likely the principal contributor to this positive index contrast, given its higher concentration in the glass. Additionally, the potential overlap of the Ce signal with Ba (due to overlapping WDS wavelengths) may cause Ce to imitate Ba's behavior.
Na, Al, and Zr are rarefied in regions of positive index contrast and accumulate in regions displaying a negative index contrast. In contrast, Ba and all lanthanides (Ce, Yb, Er) show the opposite trend. Due to the relatively higher concentration of Ba and Na, these elements exhibited the strongest signals. Fluorine is seen accumulated in the negative index change zone with no drainage from elsewhere. We believe this Z-contrast is arising from the micro-voids filled with fluorine gas similar to the observation of oxygen reported in silicate glasses~\cite{Skuja,Lancry,Richter15}.

\begin{figure}[t]
    \centering
    \includegraphics[trim={0 0 0 0},width=12 cm]{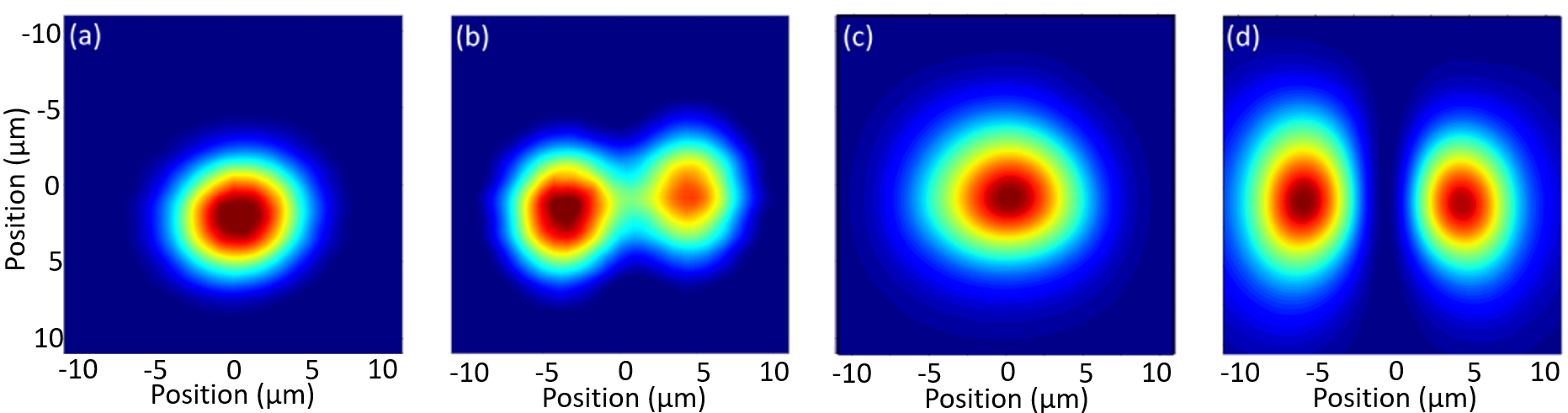}  
        \caption{(a\&b) Measured near-field profiles at  3.5~\textmu m wavelength via fibre injection for optimal alignment (a) and off-centre alignment (b). (c\&d) Simulated near-field intensity profiles (Poynting vector ($S_z$))  using the refractive index profile shown in Fig.~\ref{Rinck}(d). }
    \label{Mode}
    \end{figure}

Light from a 3.5~\textmu m laser was injected into the waveguides.  Owing to their high index contrast and the relatively wide extent of the guiding region, the first higher-order mode (LP$_{11}$) could be excited by lateral shifting of the injection fiber, as shown in Fig.~\ref{Mode} (a-b). 
The guided modes supported by the waveguide are simulated 
and found to be in good agreement with the experimental observations 
(Fig.~\ref{Mode} (c-d)). 
The simulation was done using Synopsys FemSIM\textsuperscript{\texttrademark} 
which is a mode solver software based on the Finite Element Method (FEM). 
The measured refractive index profile shown in Fig~\ref{Rinck}(b) 
is imported into the simulation after adjusting the background index to 1.4778 
for the wavelength of 3.5~\textmu m~\cite{gan1995optical}.
The entire area of the positively and negatively changed indices 
is included in the simulation, 
which has been proportionally scaled according to the dispersion relation. 
The grid size of the simulation is 200~nm both in $x$- and $y$-directions. 
The $z$-component profiles of the Poynting vector, $S_z$, 
are shown in Figs.~\ref{Mode}(c) and (d) 
to be compared with the measured intensity profiles. 
It should be noted that the fields are confined 
in the positively changed area of the waveguide. 
There was a slight discrepancy in the dimensions, with a 15\% difference along the horizontal axis and 7\% along the vertical axis. This discrepancy could be attributed to several factors, including (i) experimental uncertainties in measuring the waveguide's refractive index, (ii) uncertainties in measuring the mode profile at 3.5~\textmu m , and (iii) variations in the Sellmeier coefficients between the bulk glass and the modified regions, considering the compositional changes involved in the waveguide modification.

\begin{figure}[t]
    \centering
    \includegraphics[trim={0 0 0 0},width=6 cm]{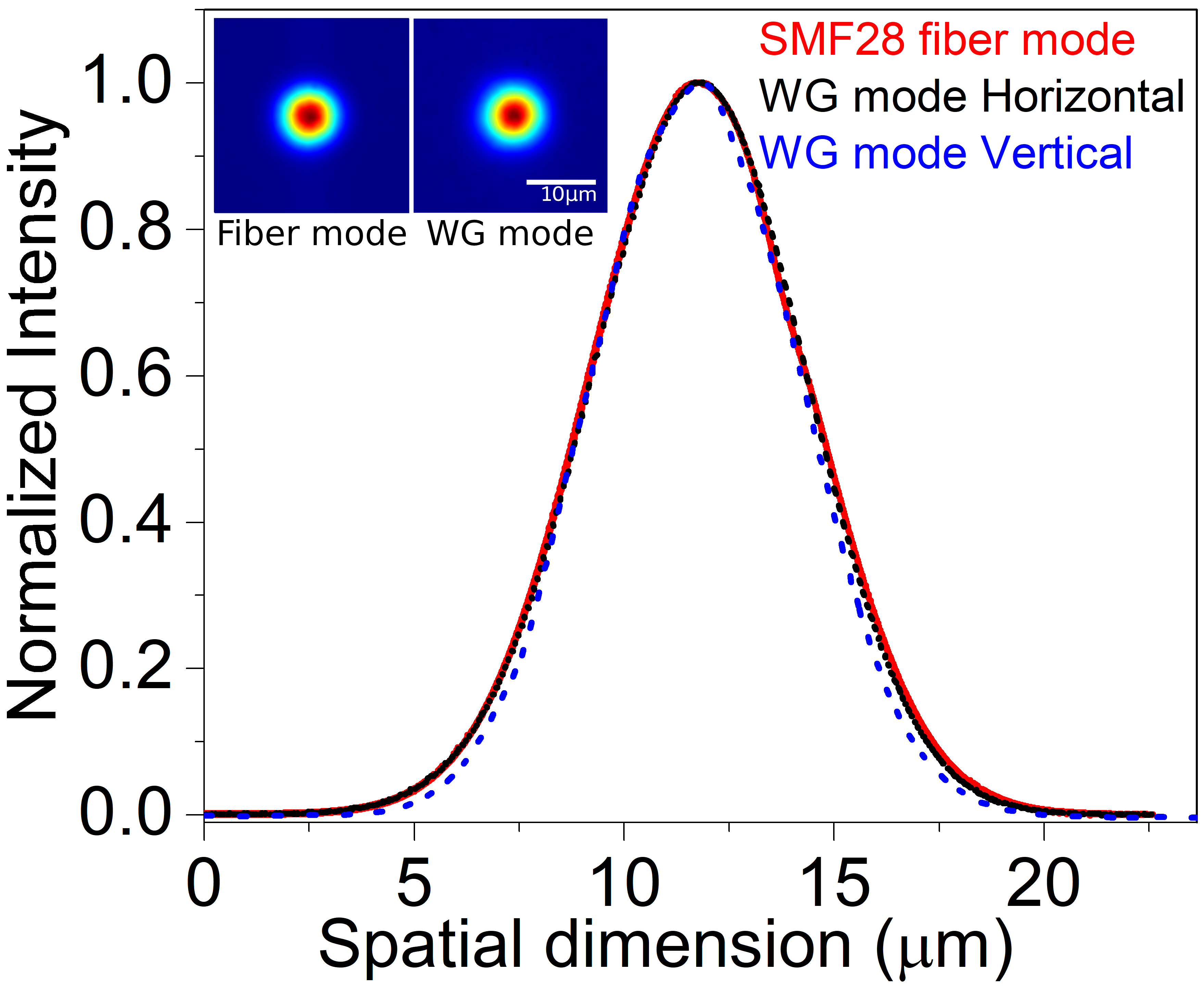}  
        \caption{Horizontal and vertical line profiles of a 1550 nm mode propagating through a 6 µm wide waveguide, fabricated using 4~\textmu J pulse energy, a writing speed of 40~\textmu m/s, and a 1 \textmu m pitch, are overlaid with the line profile of a mode from an SMF28 fiber. The inset displays the 2D mode profiles of both the fiber and the waveguide. }
    \label{1550_Modes}
    \end{figure}

The glass sample was doped with Er$^{3+}$, Yb$^{3+}$ and Ce$^{3+}$ to design a waveguide laser operating in the optical communication's C-band. Ce$^{3+}$ was doped to increase the 1.5 \textmu m erbium emission by resonance energy transfer between Er$^{3+}$ in the $^4I_{11/2}$ state and Ce$^{3+}$ in $^4F_{7/2}$ state promoting the radiative $^4I_{13/2}$ $\rightarrow$ $^4I_{15/2}$ 1.5 \textmu m emission in the former. Across the entire parameter space that was explored all exhibited multimode behavior at 3.5~\textmu m. Hence at the shorter wavelength of 1.5~\textmu m the waveguides support even more higher-order modes, scaling $\propto 1/\lambda^2$. For efficient laser operation when coupling a fiber Bragg grating (FBG) in a standard telecommunication optical fiber to the waveguide chip the mode-overlap between the fiber mode and the fundamental mode of the waveguide has to be near unity (low excitation of higher order modes) at the waveguide to fibre interface. \\
A numerical simulation was conducted using the refractive index map obtained (Fig.~\ref{Rinck}(d)), which confirmed that the waveguide width should be 6~\textmu m for single-mode operation at a 1550 nm wavelength. A detailed analysis is provided in ~\cite{fernandez2024}. The 12~mm long waveguide, written with an energy of 4~\textmu J, a speed of 40~\textmu m/s, and a pitch of 1 \textmu m, demonstrated a lowest insertion loss of 0.21 dB/cm at 1620~nm (well outside the absorption band of Er$^{3+}$ ion in fluoride glass~\cite{Miniscalco}). Index matching gel was applied to eliminate the air gap at the fiber-waveguide interface, effectively mitigating Fresnel losses. The loss measurement was carried out using a broad band ASE source and an optical spectrum analyser as described in~\cite{Beppe}. To our knowledge, this is the lowest loss waveguide in combination with a high positive index change reported to date. The 1550 nm mode profile compared to the standard single mode SMF28 fiber is provided in Fig.~\ref{1550_Modes}.

\begin{figure}[t]
    \centering
    \includegraphics[trim={0 0 0 0},width=12 cm]{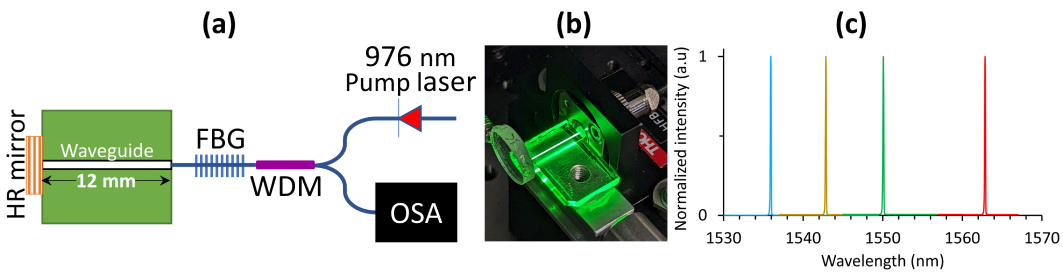}  
        \caption{(a) Schematic of the hybrid fiber/waveguide laser cavity. (b) Photograph of the actual set-up. Butt coupled fiber ferrule and the HR mirror are visible on either side of the excited waveguide. (c) Lasing output spectra obtained through the use of different FBGs of at 1535.93~nm (6.49\% output coupling, 93.51\% FBG reflectivity), 1542.85 nm (10\%), 1550.08~nm (2.5\%) and 1562.85~nm (8.92\%) centre wavelength, respectively.}
    \label{Laser}
    \end{figure}

\begin{figure}[t]
    \centering
    \includegraphics[trim={0 0 0 0},width=8 cm]{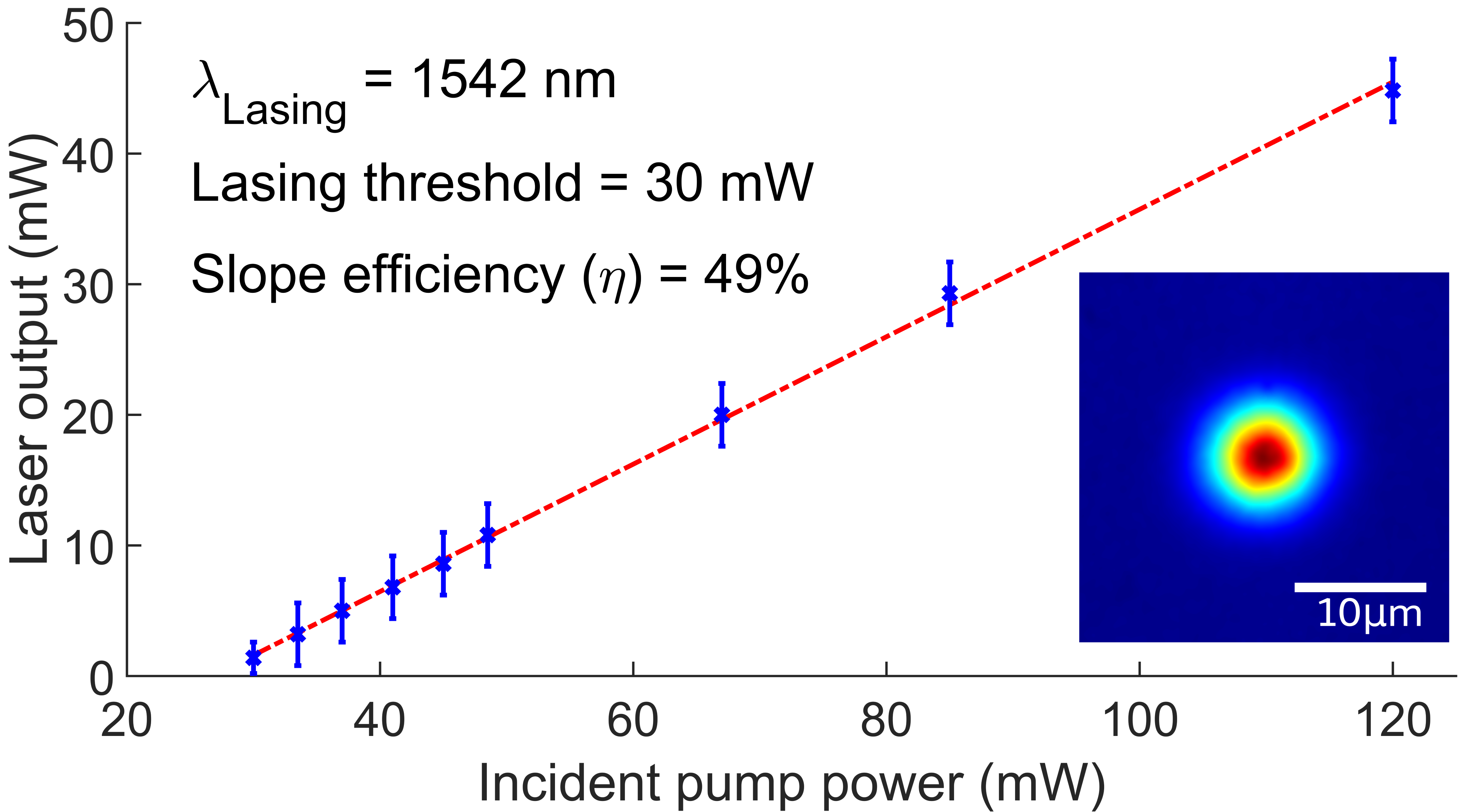}  
        \caption{Slope efficiency of the laser at 1542.85~nm and 10\% output coupling. The inset shows the near-profile of laser measured based on leakage through the HR mirror. }
    \label{Slope}
    \end{figure}

This waveguide was incorporated into a laser cavity, whose schematic is shown in Fig.\ref{Laser}(a). A high-reflective mirror is butt-coupled to one side of the waveguide and a single-mode SMF-28 fiber with Bragg grating is coupled to the other end. A fiber wavelength division multiplexer (WDM) serves to separate the laser emission at 1550~nm and pump light from a 976~nm laser diode. The actual photo of the setup is shown in Fig.\ref{Laser}(b), showing the green emission of Er$^{3+}$:$^4S_{3/2}$ $\rightarrow$ $^4I_{15/2}$ due excited state absorption under 976~nm excitation. Fig.~\ref{Laser}(c) shows the output spectrum for different FBGs. The percentage of output coupling (leakage through the FBG) for every grating central wavelength are listed in the caption of Fig.\ref{Laser}.  \\
At 1542.85~nm the threshold pump power (incident) for lasing was 30 mW, measured at the output of the FBG incident on the waveguide. The slope efficiency  with 10\% output coupling was characterized and found to be 49\%, see Fig.~\ref{Slope}. A maximum output power of 45~mW was measured after the WDM with an incident pump power of 120~mW. WDMs typically have a loss of 0.5 dB~\cite{WDM} at the measured wavelength, indicating that the maximum laser power could be recalculated as 51 mW between the FBG and WDM. This report is superior to the previous report of waveguide lasing in fluoride glass with similar rare-earth doping ~\cite{Lancaster16}. The strong laser performance in a highly confined waveguide, combined with low-loss direct coupling to optical fibers, makes this system ideal for developing advanced integrated hybrid planar-fiber frequency combs. Depending on the rare-earth dopant, these combs can operate across a broad spectral range, from the visible to mid-infrared regions~\cite{Hebert:17}
.

\section{Conclusion}
In conclusion, this study marks a significant milestone in the development of integrated optical elements in fluoride glasses for both visible and mid-infrared applications. Overcoming decades of challenges, the research has yielded optical waveguides with an unprecedented positive refractive index change, exceeding $2 \times 10^{-2}$. The investigation into the origins of the index change is attributed to material densification driven mainly by the migration of barium in the revised fluoride glass composition. \textcolor{black}{Such a strong migration to form optical waveguides eluded in the past when waveguides were inscribed in the composition which contained lanthanum fluoride.} The current recomposition ensured the migration of heavy barium atoms to positive index zone densifying with an increase in local index. The optimal fabrication conditions involved energy per pulse within the range of 2 to 6 \textmu J, feedrates spanning from 40 to 150 \textmu m/s, and a multiscan pitch of 1 \textmu m. The waveguides were approximately 12 \textmu m wide and 14 \textmu m tall, which then yielded a multimoded behaviour at 3.5 \textmu m wavelength. The losses were as low as 0.21 dB/cm which were measured at a wavelength of 1.62 \textmu m. A broad band butt coupled HR mirror and a fiber bragg grating was used to form a cavity to lase these active waveguides at 1542.83 nm wavelength.

\begin{backmatter}
\bmsection{Funding}

\bmsection{Acknowledgments}
This research was supported by the Australian Research Council Centre of Excellence in Optical Microcombs for Breakthrough Science (project number CE230100006) and funded by the Australian Government. TTF, YH and DL acknowledge support received from Electro Optic Systems Pty. Limited. This work was performed in-part at the OptoFab node of the Australian National Fabrication Facility, utilizing NCRIS and NSW state government funding. This work was performed in-part at the Adelaide Microscopy and Microscopy Australia facility funded by the National Collaborative Research Infrastructure Strategy (NCRIS) at the University of Adelaide.

\medskip

\medskip

\noindent The authors declare no conflicts of interest.

\bmsection{Data availability} Data underlying the results presented in this paper are not publicly available at this time but may be obtained from the authors upon reasonable request.

\end{backmatter}

\bibliography{References}

\end{document}